**β-Carotene bioavailability and conversion efficiency are significantly affected by sex in rats.** *First observation suggesting a possible hormetic regulation of vitamin A metabolism in female rats.*

Patrick BOREL[1*], Romane TROADEC[1**], Morgane DAMIANI[1], Charlotte HALIMI[1], Marion NOWICKI[1], Philippe GUICHARD[1], Marielle MARGIER[1], Julien ASTIER[1], Michel GRINO[1], Emmanuelle REBOUL[1], Jean-François LANDRIER[1]

[1]C2VN, INRAE, INSERM, Aix-Marseille Univ, Marseille, France.

*Corresponding author: Patrick.Borel@univ-amu.fr

UMR INRAE/INSERM/AMU "Center for CardioVascular and Nutrition Research of Marseille"

Faculté de Médecine ; 27, boulevard Jean Moulin ; 13005 Marseille, France

Phone: +33 (0)4 91 32 42 77

**co-first author.

**Abbreviations:** area under the curve (AUC); Beta-carotene oxygenase 1 and 2 (BCO1 and BCO2); high vitamin A diet (HVA); intestine specific homeobox (ISX); lecithin retinol acyltransferase (LRAT); low vitamin A diet (LVA); medium vitamin A diet (MVA); retinoic acid receptor (RAR); retinoic X receptor (RXR); retinol binding protein 4 (RBP4); retinol binding protein 4 receptor 2 (Rbpr2); Signaling receptor and transporter of retinol (STRA6); transthyretin (TTR); vitamin A (VA);

**Abstract**

**Scope**: To study the effect of variation in dietary vitamin A (VA) content on its hepatic and intestinal metabolism. **Methods and results**: Adult female and male rats were fed with diets containing 400, 2300, or 9858 IU/kg VA for 31-33 weeks. VA concentrations were measured in plasma and liver. Bioavailability and intestinal conversion efficiency of β-carotene to VA were assessed by measuring postprandial plasma β-carotene and retinyl palmitate concentrations after force-feeding rats with β-carotene. Expression of genes involved in VA metabolism, together with concentrations of RBP4, BCO1 and SR-BI proteins, were measured in the intestine and liver of female rats. Plasma retinol concentrations were lower and hepatic free retinol concentrations were higher in females than in males. There was no effect of dietary VA content on β-carotene bioavailability and its conversion efficiency, but bioavailability was higher and conversion efficiency was lower in females than in males. The expression of most genes exhibited a U-shaped dose response curve depending on VA intake. **Main conclusions**: β-Carotene bioavailability and conversion efficiency to VA are affected by the sex of rats. Results of gene expression suggest a hormetic regulation of VA metabolism in female rats.

## 1) Introduction

Vitamin A (VA), i.e. all-*trans* retinol and its derivatives, is involved in numerous biological functions in humans and other mammals, the best known of which is its role in vision, via photochemical isomerization of 11-*cis*-retinal, [1] and in the modulation of gene expression, via binding of retinoic acid to the retinoic acid receptor (RAR) and retinoic X receptor (RXR) transcription factors. [2-4] These functions require tight regulation of the metabolism of VA and of its precursors, i.e. provitamin A carotenoids (mainly β-carotene, α-carotene and β-cryptoxanthin). These regulations are ensured by numerous proteins more or less specific to VA or provitamin A or to their different metabolites. [5] For instance, retinol binding protein 4 (RBP4), [6] together with transthyretin (TTR), are responsible for the blood transport of retinol. [7, 8] Beta-carotene oxygenase 1 and 2 (BCO1 and BCO2), which are expressed in several tissue including the gastrointestinal tract and the liver, [9, 10] allow mammals to synthesize retinoids from provitamin A carotenoids. [11, 12] Signaling receptor and transporter of retinol (STRA6) and retinol binding protein 4 receptor 2 (Rbpr2) are responsible for the uptake of retinol from the blood by peripheral tissues. [13, 14] Most of these proteins, except STRA6, are synthesized in the intestine, which is the gateway to dietary VA and provitamin A in our body, and in the liver, which is the main storage site of VA (about 70% of VA body stores), playing a central role in regulating VA delivery to the body. [5, 15, 16]

Daily VA and provitamin A intakes are very variable, due to the high variability of the concentrations of the different VA species in food [17]. Indeed, there is practically no preformed VA, i.e. retinol and retinyl esters, in food of plant origin, whereas provitamin A carotenoids are mainly found in these foods. Furthermore, some foods, e.g. liver from livestock, can provide a very large amount of VA in a single meal [18]. The sudden entry into the body of large quantities of a molecule with such different biological effects certainly explains why evolution has selected mechanisms that allow a very fine regulation of the concentrations of VA and its metabolites in the blood and in the different tissues [5]. This regulation requires a very fine control of the synthesis of proteins involved in VA metabolism. [16] It would seem logical that this regulation is sensitive to the VA status of the organism, and mediated by retinoic acid. This hypothesis is supported by the work of Vanvliet et al. who found that BCO1 activity was higher in VA deficient rats than in rats with high VA intake. [19] The existence of a diet-responsive regulatory network that controls β-carotene absorption and VA production through negative feedback regulation of Bco1 and scavenger receptor class B member 1 (Scarb1) expression via intestine specific homeobox (ISX) in the intestine was later suggested. [20-22] Furthermore, it was recently shown that the catalytic activity of lecithin retinol acyltransferase (LRAT) coordinates this regulation. [23] However, Goswami et al. [24] did not observe an effect of the VA status on β-carotene absorption and conversion to VA in rats. Furthermore, Lemke et al. [25] observed a lower conversion efficiency of β-carotene after VA supplementation in 2 women, which agrees with the regulatory network, together with a higher absorption efficiency of β-carotene in these volunteers, which apparently disagree with this network. In conclusion, this regulation is not found in all studies and it is therefore one of the objectives of this study to verify the existence of this regulation following a variation in the VA content of the diet in male and female rats.

Strikingly, all the animal studies which have been devoted to study the effect of the VA content of the diet on the metabolism of this vitamin (16 studies, 80% of which were carried out in rodents, supplemental table 1), were performed on male animals only. Paradoxically, the only study on the opposite sex has been carried out in the human species and on two women only. [25] However, it has been known for many years that there are differences in the metabolism of VA between the two sexes [26, 27]. Indeed, it has been shown that retinolemia is lower in females than in males, both in rodents [26, 28] and in humans [27]. Conversely, it was shown that, in rodents, the concentration of VA in the liver was higher in females than in males [26, 28, 29]. These differences in VA metabolism between males and females are also suggested by other observations: females have higher CRBP concentrations than males in certain tissues [30], the effect of β-carotene supplementation is sex-dependent [31] and there is a female-specific mechanism for retinoic acid generation in some tissues [32]. We hypothesize that there may be other differences between males and females with respect to VA metabolism, such as differences in the bioavailability of β-carotene and in the efficiency of its conversion to VA. Indeed, such differences might explain the different effects observed between males and females in supplementation studies aimed at correcting deficiencies or low VA status. [33, 34]

The objective of this study was to deepen our knowledge on the effect of the VA content of the diet on VA metabolism considering the effect of sex. For that purpose, we submitted groups of rats to diets with different levels of VA. At the end of the dietary intervention, we measured hepatic and plasma concentrations of VA as well as expression of genes and proteins involved in VA metabolism. We also forced-fed the rats with β-carotene to measure its bioavailability and its conversion efficiency to VA.

## 2) Material and methods

*2.1 Chemicals.*

Ethanol, *n*-hexane, isopropanol, trichloromethane and HPLC grade dichloromethane, methanol, acetonitrile and water were purchased from Carlo Erba reagents (Val de Reuil, France). β-Carotene (HPLC purity > 96%) was from Carotenature GmbH (Münsingen, Switzerland). Retinol, retinyl palmitate, tocol, triolein, phospholipids, NaCl, sodium citrate, Tris hydroxide, bovine serum albumin and protease inhibitor cocktail were from Sigma Aldrich (Saint-Quentin Fallavier, France). Sevoflurane was from Baxter (Lublin, Poland), TRIzol reagent was from Euromedex (Souffelweyersheim, France). Dithiothreitol was from Thermo Fischer Scientific (Les Ulis, France). Phosphate buffered saline (PBS) was from life Technologies (Illkirch, France).

*2.2 Raising rats on diets that contained different concentrations of VA.*

We used rats rather than mice to correctly measure postprandial VA responses following gavages with β-carotene. It is indeed not possible to collect sufficient blood volumes, at least 500 μL, at each postprandial point in mice. All experimental procedures were approved by the local animal care and use committee (agreement number D 13-055-20). Three groups of 10-week-old Sprague Dawley RjHan:SD rats (Janvier Labs, Le Genest-Saint-Isle, France) were housed under standard conditions of light (12-h light/dark cycle; lights on at 08h00 am) and temperature (22-24°C) with free access to tap water and breeding pellet diet. Each group of rats consisted of 8 males and 12 females. A first group was put on a diet that contained 400 IU/kg VA (AIN-93G W/ 400 IU/KG VITAMIN A), thereafter called low VA diet (LVA), a second group was put on a diet that contained 2300 IU/kg VA (AIN-93G W/ 2300 IU/KG VITAMIN A), thereafter called medium VA diet (MVA), and a third group was put on a diet that contained 9858 IU/kg VA (AIN-93G W/ 9858 IU/KG VITAMIN A), thereafter called high VA diet (HVA). These diets were developed by Test Diet Limited (London, UK) and were provided to animals for 31-33 weeks.

2.3 Dosage information / Dosage regimen

The effects of dietary VA concentrations and sex of rats on β-carotene bioavailability and conversion efficiency into VA were evaluated by force-feeding the rats with β-carotene incorporated in a lipid emulsion. A single dose of 3 mg β-carotene was administered at fast to half of the rats in each group. This provided about 4.7 mg/kg bodyweight to males and about 8.8 mg/kg bodyweight to females. This is equivalent to 0.38-0.71 mg/kg bodyweight in humans, which is only achievable through supplements. The dose of β-carotene was chosen following a preliminary experiment which enabled us to determine the amount of β-carotene necessary to correctly measure this molecule and its main metabolite, i.e. retinyl palmitate, secreted in the plasma during the post-gavage period. Knowing that rats convert β-carotene to VA very efficiently, this preliminary experiment showed that at least 3 mg of β-carotene/gavage were required to allow its detection in the plasma during the postprandial period.

β-Carotene was first mixed with 0.33 mL triolein and placed in glass tubes that were stored at -80°C. Emulsions were prepared on the day of the force-feeding, as follows: 0.66 mL of an aqueous solution containing 0.9% NaCl, 2% bovine serum albumin and 0.1% phospholipids was first deposited on the triolein + β-carotene mixture. Then, oil-in-water emulsions were created by vortexing the mixture for 6 min then sonicating it, with a sonication probe, for 2 consecutive sessions of 20 s. This procedure led to oil-in-water emulsions that remained relatively stable for the duration of the gavage experiment (approximately 2 h).

*2.3 Gavage experiments to assess the bioavailability of β-carotene and its conversion efficiency into VA in the intestine.*

Four males and six females in each group were force-fed. A blood sample (about 500 µL) was first collected from a tail nick into tubes that contained 50 µl of a 0.109 M sodium citrate solution (baseline). The force-feeding was performed in overnight fasted rats after a mild anesthesia using sevoflurane. The procedure was as follows: the emulsion containing β-carotene was drawn into a syringe and injected into the stomach of the rats with a nasogastric tube. This was considered as time 0 of the postprandial experiment. Blood samples (about 500 µL) were then taken from the tail of the rats 1.5 h, 3 h, 4.5 h and 6 h after gavage. The last sample was obtained by cardiac puncture under deep sevoflurane anesthesia using a syringe filled with 0.109 M sodium citrate (1/10 of the blood volume to be sampled). Blood was immediately centrifuged at 4,000 x *g* for 10 min at 4°C and the resulting plasma was immediately frozen at – 80°C.

The animals that were not force-fed, i.e. 4 males and 6 females of each group, were euthanized after overnight fasting, to collect blood and tissues of interest (liver and intestine).

*2.4 Collecting hepatic and intestinal samples*

Small intestine was quickly harvested after the death of the animals and carefully rinsed with ice-cold PBS. Eight cm of intestine were carefully harvested in the duodenal, jejunal and ileal part of the intestine and aliquoted in 2 samples. Both segments of the intestine were opened longitudinally and the intestinal mucosa was scraped off. One sample was kept for qPCR experiments, the second one for protein measurements. Liver was weighed and the lower left quarter of the main lobe was taken. All tissue samples were immediately, frozen in liquid nitrogen and stored at -80°C.

*2.5 Measurement of β-carotene and VA concentrations in plasma and hepatic samples*

The molecules of interest had first to be extracted in an organic phase. This first step was performed at room temperature, avoiding direct exposure to light. Concerning plasma samples, approximately but precisely 250 μL were used. The volumes were adjusted to 500 μL with distilled water. Then, 500 μL of an ethanolic solution containing around 0.5 μg/mL tocol (used as internal standard to calculate extraction yield) were added. Concerning hepatic samples, approximately but precisely 100 mg of liver were weighted and ground 5 min at 30 rotations/min in 1 mL of PBS. Then, 500 μL of an ethanolic solution containing around 0.5 μg/mL tocol were added. After this step of protein precipitation by ethanol, 2 mL of hexane were added in each tube, followed by a 10 min stirring using an orbital shaker, and a 10 min centrifugation at 1225 x *g* at 4°C. The hexane upper phase was collected. The remaining inferior phase was re-suspended with 2 mL of hexane, followed by the same steps than previously. Again, the upper phase was collected and added to the first one. Tubes were left to evaporate under nitrogen until obtaining a dried extract. All dried extracts were dissolved in 200 μL HPLC mobile phase (see below). A volume of 50–180 μL was used for HPLC analysis. The molecules of interest were analyzed by HPLC using a 250 x 4.6 mm RP C18, 5 μm Zorbax column (Agilent Technologies, Santa Clara, CA, USA) preceded by a guard column and maintained at a constant temperature (35°C). The mobile phase consisted of acetonitrile-dichloromethane-methanol (70/20/10; vol/vol/vol). The flow rate was 1.8 mL/min.[35] The HPLC system comprised a Shimadzu separation module and a photodiode array detector. Compounds were detected at their maximum absorption wavelengths, i.e. 265 nm for tocol, and 325 nm for retinol and retinyl esters (we measured retinyl palmitate, retinyl stearate, retinyl oleate and retinyl linoleate, which are the main retinyl esters present in rat hepatic stellate cells[36]) and 450 nm for β-carotene. Retinol, β-carotene and retinyl palmitate were identified by retention times and absorption spectra coincident with authentic (>95% pure) standards. Retinyl linoleate, retinyl oleate and retinyl stearate were identified by spectral analysis and quantified regarding their molar extinction coefficient ratio compared to retinyl palmitate. Quantification was performed using Chromeleon software (version 6.50 SP4 Build 1000) comparing peak area with standard reference curves.

*2.6 Measurement of gene expression in hepatic and duodenum samples.*

These measurements were performed only in females. Total RNA was extracted from the liver and intestinal mucosa using TRIzol reagent as previously described. [37, 38] Total RNA (1 µg) was used to synthetize cDNAs using random primers and Moloney murine leukemia virus reverse transcriptase (Thermo, Courtaboeuf, France). Real-time quantitative PCR analyses were performed using a Light Cycler 480 system (Roche, Basel, Switzerland). For each condition, gene expression was quantified in duplicate using the primers listed below, and 18S rRNA was used as the endogenous control in the comparative cycle threshold (CT) method. [39]

| Gene | Sense primer | Reverse primer |
| --- | --- | --- |
| *Bco1* | TTGCTGCTGAAGTACAATGGA | GAGGACTTGGTGAGAACGTCA |
| *Bco2* | TTGACCTGTGCTGTGAGGAT | TGCCTTTGCCTTATAGACCTG |
| *Cd36* | TGAGAAGTCTCGAACACTGAGG | TCCAAACACAGCATAGATGGAC |
| *Ces1e* | AGAGCCCTGGAGCTTCGTA | GGGGTCTTGGGAGCACAT |
| *Dgat1* | TACGGCGGGTTCTTGAGAT | CGTGAATAGTCCATGTCCTTGA |
| *Dgat2* | GTGTGGCGCTATTTTCGAG | GGTCAGCAGGTTGTGTGTCTT |
| *Isx* | GAAGCCCACACAGGATCAG | AGTGGGTGAAGTGGAAGAGC |
| *Lrat* | GCAGATACGGCTCTCCTATCA | CGAATGAGTATCTTCACAGTCTCG |
| *Pnpla3* | CACCAACCTCAGTCTTCGTCT | AAGCACAGCTCTCCCATCAC |
| *Rbp1* | AGGCATAGATGACCGCAAGT | CACTGAAGCTTGTCACCATCC |

| *Rbp2*  | AAGGGGAGAAGGAGAATCGT | GTCACCGCAGGTCAGCTC |
| *Rbp4*  | GTGAGGCAGCGACAGGAG   | TTGAGGGTCTGCTTTGACAG |
| *Scarb1* | GGTGCCCATCATTTACCAAC | GCGAGCCCTTTTTACTACCA |
| *Ttr*   | CATTCCATGAATACGCAGAGG | TGTAGTGGCGATGACCAGAG |
| *18S*   | CGCCGCTAGAGGTGAAATTCT | CATTCTTGGCAAATGCTTTCG |

*2.7 Quantification of proteins of interest in the liver and intestine.*

BCO1, SR-BI and RBP4 proteins were quantified in hepatic and intestinal samples of the female rats using commercial ELISA kits. Pieces of around 3 mm$^3$ of liver or intestinal mucosa were first mixed with PBS and a protease inhibitor cocktail (Sigma Aldrich). Then, they were homogenized by grinding and sonication. The total protein concentration of homogenates was quantified using BCA protein assay kit (Thermo Fisher Scientific). BCO1 was quantified from 60 μg of liver homogenate and 20 μg intestinal homogenate using a colorimetric immunometric enzyme immunoassay kit (Rat β,β-carotene 15,15-Dioxygenase (BCO1) ELISA Kit, Abbexa Ltd, Cambridge, UK). RBP4 was quantified from 1μg of liver homogenate using immunoassay kit (Rat Retinol Binding Protein 4, Plasma (RBP4) ELISA Kit, Dldevelop, Shuigoutou, Wuxi/Jiangsu, China). SR-BI was quantified from 20 μg intestinal homogenate using immunoassay kit (Scavenger Receptor Class B Member 1 (SRB1) ELISA Kit, Cell Biolabs Inc., San Diego, CA USA).

*2.8 Methods for calculating the postprandial responses in VA and scientific bases on which the bioavailability of β-carotene and its conversion rate in VA were estimated.*

The postprandial responses of the different forms of VA measured in the plasma following the gavages with β-carotene, i.e. β-carotene, retinyl palmitate and retinol, were estimated by calculating the incremental areas under the curves of plasma concentrations of these molecules as a function of time. These calculations yielded areas under the curves (AUC) which were used to compare the responses measured in each group of rats. [40]

According to what is known about the metabolism of β-carotene and VA in mammals, [5] β-carotene found in the blood in the postprandial period in a rat that did not have any β-carotene in the diet before force-feeding with β-carotene necessarily comes from newly absorbed β-carotene. In addition, postprandial blood retinyl esters, mainly retinyl palmitate, [41] necessarily come from the intestinal conversion of newly absorbed β-carotene into retinal, retinol then retinyl esters, mainly retinyl palmitate. [41] Finally, concerning postprandial blood retinol, it is assumed that it comes almost exclusively from the liver that secretes it in the blood in association with RBP4 and TTR.

Therefore, the sum of postprandial β-carotene and retinyl ester responses, i.e. their AUC, after force-feeding with β-carotene gives a good estimate of the total amount of β-carotene that was absorbed, i.e. that was bioavailable. Concerning the efficiency of the intestinal conversion of β-carotene into VA, based on what is assumed regarding the intestinal metabolism of β-carotene, it was estimated according to the following formula: % conversion = AUC retinyl ester / (AUC β-carotene + AUC retinyl ester) x 100.

*2.9 Statistical analysis.*

Results are expressed as means ± SEM (standard error of the mean). Two-way ANOVA, with diet and sex as factors, or one-way ANOVA, with diet as factor, were used to assess the effect of diet or sex, or both, on each measured variable. When a significant ($p < 0.05$) effect was detected, post-hoc Tukey-Kramer tests were used to compare means from the different groups. Statistical analyses were performed using StatView software version 5.0 (SAS Institute, Cary, NC, USA). In all cases, $p < 0.05$ was considered statistically significant.

## 3) Results

*3.1 Effect of dietary VA content on plasma and hepatic concentrations of VA.*

Retinol plasma concentrations were measured in the different groups of rats (subgroups that were not force-fed) at fast, i.e. after consuming diets providing different concentrations of VA for 31-33 weeks. **Figure 1** and two-way ANOVA with diet and sex as factors clearly show that there was no effect of diet ($p = 0.389$) but a significant effect of sex ($p < 0.0001$) on plasma retinolemia. Males had on average about 2.3 times higher retinolemia than females.

Concerning hepatic VA, both retinol and several retinyl esters were measured (**Figures 2 and 3**). The first observation (**Figure 2A**) is that the total amount of hepatic VA, i.e. the sum of the concentrations of retinol and retinyl esters increased proportionally to the concentration of VA in the diet, in both females and males (diet effect $p < 0.0001$ and the sex effect was marginally significant, i.e. $p = 0.064$). In fact, there was a strong correlation between the total amount of hepatic VA and the VA content of the diet, e.g. in females the concentration of hepatic VA was equal to 0.189 x the concentration of VA in the diet (IU/kg), coefficient of determination $R^2 = .997$ (Data not shown). The second observation that emerges from **Figure 2** is that retinyl esters (**Figure 2B**) represents the main hepatic VA form, in all groups. Furthermore, the mean proportion of VA as free retinol was significantly higher in females than in males, representing about 10% and 3% of total VA, respectively. The third and noteworthy observation was that there was a significant effect of sex ($p = 0.0092$), diet ($p = 0.0011$), and their interaction ($p = 0.0015$) on hepatic free retinol concentrations. Concerning the different species of retinyl esters (**Table 1**), the main one was retinyl palmitate followed by stearate, oleate and linoleate. The higher the VA content of the diet, the higher the concentration of each of the retinyl esters. The differences in concentration between males and females increased as the VA content of the diet increased. It is also very interesting to observe that the proportion of retinyl palmitate was always higher in males than in females and that it was the reverse for retinyl stearate.

*3.2 Effect of the VA content of the diet on postprandial plasma β-carotene and VA concentrations following force-feeding with β-carotene.*

The effect of the VA content of the diet, and consequently of the rat VA status, on both β-carotene bioavailability and on its intestinal conversion efficiency into VA was evaluated by measuring postprandial plasma concentrations of retinol, retinyl palmitate and β-carotene after force-feeding rats with a high dose of β-carotene. As expected, and unlike retinol, neither β-carotene nor retinyl palmitate were detected in plasma prior to gavage (data not shown). **Figure 3A** shows that the retinol incremental AUCs, corresponding to retinol responses, were close to 0 for the MVA and HVA groups, while they were clearly positive for the LVA groups. Due to the great variability of the LVA male results, the diet effect was at the limit of significance ($p = 0.0706$). The β-carotene responses (**Figure 3B**) were dramatically higher in females than in males (about 10 times higher; $p < 0.0001$). Conversely, the β-carotene response was not significantly affected by diet, although it was interesting to mention that there was a tendency for a lower response in the MVA female group as compared to the two other female groups. Finally, with regard to retinyl palmitate (**Figure 3C**), there was no significant effect of diet and of sex on this phenotype, although again there was a tendency for a higher response in the MVA female group as compared to the two other female groups.

**Figure 4** shows the effect of the dietary VA content on the bioavailability of β-carotene and on its conversion efficiency to VA (see material and method section where it is explained how these variables were calculated). β-Carotene bioavailability (**Figure 4A**) was significantly lower in males than in females (about 55% on average; $p = 0.0002$) but it was not affected by the diet, in both males and females. Concerning conversion efficiency (**Figure 4B**), males had a significantly higher conversion rate than females ($p < 0.0001$), about 94% on average. Moreover, while the diet did not affect the conversion efficiency in males, it significantly affected it in females ($p = 0.0086$). More precisely, females in the MVA group had a higher conversion efficiency than females of the two other groups: + 51% (not significant) as compared to LVA females, and + 252% ($p < 0.05$) as compared to HVA females.

*3.3 Effect of the VA content of the diet on the expression of genes involved in VA metabolism*

The expression of several genes coding for proteins known to be involved in hepatic metabolism of VA [5] was measured in the 3 groups of females (subgroups that were not force-fed). Results are presented in **Figure 5A**. Strikingly, all genes, except *Rbp2* and *Pnpla3,* were more expressed in the LVA and HVA groups compared to the MVA group. Furthermore, for most, there was no-significant difference between the LVA and the HVA groups.

Results on the expression of genes coding for proteins known to be involved in the metabolism of VA in the duodenum are presented in **Figure 5B**. It is first of all important to observe that the variations in gene expression are markedly lower than in the liver, by a factor of approximately 10. Similarly, to what is observed in the liver, duodenal gene expressions were roughly equivalent and higher in the LVA and HVA groups compared to the MVA groups, with the notable exception of *Isx*, whose expression correlates positively, and almost significantly ($p = 0.084$), with the concentration of VA in the diet (Data not shown). The fact that only the differences in expression of *Dgat1* are significant is likely explained by the fact that the amplitudes of the variations in gene expression induced by the diet were much lower than in the liver and that the number of samples in each group was not sufficient for the differences to be significant.

*3.4 Effect of the VA content of the diet on the concentration of proteins involved in VA metabolism.*

The concentrations of two hepatic proteins whose gene expression was significantly modulated by the VA content of the diet, namely BCO1 and RBP4, were measured by ELISA. The results, presented in **Figure 6A**, show that these two proteins were present in greater concentrations in LVA and HVA groups as compared to the MVA group. In addition, the concentrations measured in the LVA and HVA groups were not significantly different.

Two proteins involved in intestinal metabolism of VA, i.e. BCO1 and SR-BI, were quantified in the duodenum, jejunum and ileum of the female rats. **Figures 6B** and **6C** show that, in all cases, the concentrations of these two proteins were greater in the LVA and HVA groups as compared to the MVA group although the differences are not always significant, probably due to lack of statistical power. But what is important to observe is that, overall the concentrations of these proteins varied according to a U-shaped dose response depending on the concentration of VA in the diet.

## 4) Discussion

The main objective of this study was to assess the effect of the VA content of the diet, and therefore of the resulting VA status, on its metabolism. Our first concern was thus to determine how VA status was modulated by the VA content of the diet. To this end, three biomarkers of VA status were measured: i) plasma retinol concentrations, ii) hepatic retinol and retinyl ester concentrations, and iii) changes in plasma retinol concentrations during the postprandial period after force-feeding with β-carotene.

The first marker, i.e. retinolemia, was not significantly different between same-sex groups that consumed diets with different concentrations of VA. Such observation was consistent with the fact that retinolemia is tightly regulated and significantly decreased only when the hepatic stores of VA are practically exhausted. [42] In agreement with previous observations, plasma retinol concentrations in all groups of females were approximately two times lower than those observed in all groups of males [26-29]. Since females had significantly higher hepatic free retinol concentrations than males, we hypothesize that hepatic free retinol was less secreted in bloodstream in females than in males. Interestingly, lower blood retinol concentrations were also observed in women as compared to men, [27, 43] This could be due to a lower hepatic RBP4 expression leading to lower blood concentrations of RBP4, as has been observed in mice. [44]

The second biomarker of VA status, i.e. total hepatic VA concentration, is acknowledged as the best biomarker of VA status. [42] This biomarker correlated positively and significantly with dietary VA concentrations, both in males and in females, showing that our nutritional intervention was successful and that hepatic accumulation of VA was linear in the studied range of dietary VA concentrations. Regarding the effect of sex on this biomarker, several observations show that VA status is modulated by the VA content of the diet differently in females vs males. The first observation, which is in perfect agreement with previous studies [26, 28, 29], is that females accumulated more VA in the liver than males when the diet was rich in VA. The second is that females always had higher concentrations of free retinol in the liver than males, which, to the best of our knowledge, has never been observed. Finally, there was a differential esterification of retinol in the two sexes since we observed, for the first time, that there was always a higher proportion of retinyl palmitate and a lower proportion of retinyl stearate in males as compared to females.

Regarding the third biomarker, it can be compared to a relative dose response (RDR) test, which has been developed for assessing VA status in human. [42, 45] The principle is as follows: a high dose of VA is given to a fasting subject suspected of being VA deficient. Retinolemia is measured at fast and 5 h after consumption of the high dose of VA. [46] If the subject is VA-deficient, retinolemia increases significantly at 5 h as compared to the value measured at fast. Following previous observations in VA depleted hepatocytes [47] and in VA deficient rats, [48] it is assumed that this is due to the sudden mobilization of a pool of apo-RBP4 which is accumulated in the liver to be released rapidly in the blood when VA becomes available. In our study, we observed that the postprandial retinol response (retinol AUC) to the β-carotene gavage was positive in the LVA groups, while it was around zero in the MVA and HVA groups, suggesting that the LVA rats had marginal VA deficiency.

Altogether, the three biomarkers show that the VA status was: 1) significantly different between groups who consumed diets with different concentrations of VA, 2) positively correlated with the VA content of the diet, 3) higher in HVA females than in HVA males, 4) deficient for the LVA groups.

After having verified that our nutritional intervention had indeed modified the VA status as desired, we can interpret the effects of this status on the metabolism of this vitamin. Regarding intestinal metabolism of VA, we first aimed to assess whether VA status affects β-carotene bioavailability and β-carotene conversion efficiency into VA. For this we interpreted the postprandial responses in β-carotene and retinyl palmitate following force-feeding with β-carotene. Concerning the results obtained in males, neither β-carotene bioavailability nor its conversion efficiency was significantly affected by VA status. This was surprising because this does not agree with the results of previous studies [19, 49] and this is apparently in disagreement with the diet-responsive regulatory network which assumes that VA deficiency increases β-carotene absorption and β-carotene conversion to VA. [22, 50] To explain this apparent discrepancy, we suggest that the VA status of the LVA group was perhaps not deficient enough to activate this regulatory network. Indeed, although the retinol responses in this group suggest VA deficiency, there were still retinyl esters in the liver of these rats. In addition, note that other authors have not observed this regulation either. [24, 25] Thus, we hypothesize that other(s) factor(s) which are different in these different studies, likely play a role in this regulation. [51] Concerning the effect of VA status on intestinal metabolism of VA in females, it is interesting to note that, as observed for gene expression, there were apparent U-shaped responses regarding the effect of VA status on β-carotene and retinyl palmitate responses. More precisely, MVA females had a lower β-carotene response together with a higher retinyl palmitate response as compared to females of the two other groups (**Figure 3**). Consequently, this led to a higher conversion efficiency of β-carotene in MVA females as compared to LVA and HVA females. This observation is in agreement with the previously reported increase in β-carotene bioavailability, associated with a decrease in its conversion efficiency into VA, following VA supplementation in 2 women. [25] The increase, or at least the non-decrease, of the bioavailability of β-carotene following VA supplementation in females is also in apparent contradiction with the regulatory network, [22, 50] but it should be remembered that all the studies leading to the suggestion of this network have been done in males and it is possible that this network does not work in the same way in females.

Conversely to the effect of the VA content of the diet on intestinal metabolism of VA, the sex effect on this metabolism was very marked in rats. Indeed, β-carotene bioavailability in males was about half of that measured in females, regardless of the VA content of the diet. Since males also converted β-carotene to VA better than females, independently of the VA status, this likely explains the much lower β-carotene response in males than in females. To our knowledge, this is the first time that such sex effect is observed on intestinal β-carotene metabolism.

Regarding the effect of VA status and sex of animals on hepatic metabolism of VA, two parameters were measured: concentrations of free retinol and concentrations of the main retinyl ester species. This allowed us to calculate and to compare total hepatic VA concentrations and relative proportions of the different VA species in the different groups. Concerning the effect of the VA content of the diet on hepatic concentrations of VA, this has been discussed above when these concentrations were used as biomarkers of VA status. Concerning the effect of the sex of animals, the higher total hepatic VA concentration in HVA females, compared to HVA males, confirms what had already been observed, [26, 28, 29] i.e. that females have a higher VA storage capacity than males. With regard to hepatic free retinol, the fact that its concentration was always higher in females than in males was notable. We hypothesize that this is due to less efficient secretion of hepatic free retinol in the blood in females. Indeed, this hypothesis agrees with the lower retinolemia in females and could be due to either a lower hepatic expression of its transporter protein, i.e. RBP4, [44] leading to a lower plasma concentration of VA, or to a difference in expression of *Cd36* between males and females. [52] Indeed, it has been shown that this protein plays a role in the mobilization of hepatic VA stores, since *Cd36*$^{-/-}$ mice accumulated higher hepatic VA stores than wild type mice. [53] The last parameter of hepatic metabolism of VA that was affected by the sex of animals was the relative proportions of each retinyl ester species. Indeed, males had higher proportions of retinyl palmitate and lower proportions of retinyl stearate than females, whatever the diet. We suggest that this is due to a difference in fatty acid composition of hepatic phospholipids between males and females. Indeed, LRAT uses the fatty acids of phospholipids to esterify retinol, and male rats have been shown to have higher concentrations of palmitate and lower concentrations of stearate in phospholipids than female rats [54].

Regarding the effect of the VA content of the diet, and thus of VA status, on hepatic gene expression, which could only be measured in female rats for various reasons, we observed that the expression of almost all genes was significantly modulated by the VA content of the diet following a U-shaped dose response. Indeed, on the whole, the gene expressions were not different in the LVA and HVA groups and significantly higher in these groups compared to the MVA group. We can reasonably exclude that these variations in gene expression were due to hepatic toxicity of VA because the average liver weight of the rats, which was different between males and females (22 g vs 12 g), was not significantly between rats of the same sex that consumed different amounts of VA (data not shown). Furthermore, the effect of the VA content of the diet on the expression of the duodenal genes showed similar U-shaped dose responses, with the notable exception of *Isx*. Although the differences between the groups were much smaller than those observed in the liver and most of them, except those of *Dgat1*, were not significant. As explained previously, to ensure the validity of the results on gene expression, we measured the effect of the VA content of the diet on the concentrations of some key proteins involved in VA metabolism in the liver (BCO1 and RBP4) and in the intestine (BCO1 and SR-BI). Although we are aware of the limitations of Elisa's BCO1 assay, the results of the Elisa for this protein, as well as those of the Elisa for the other two proteins are fully consistent with the results of gene expression. Indeed, the LVA and HVA diets led to similar protein concentrations, which were generally significantly higher than the concentrations observed with the MVA diet. In other words, the concentrations of these proteins varied according to a U-shaped dose response depending on dietary VA concentration. This is the first time, to our knowledge, that a U-shaped dose response to dietary VA is observed for expression of genes involved in VA metabolism. This seems to disagree with data showing that the expression of hepatic *Lrat* was positively correlated with VA status, [55] but we hypothesize that this is because the VA status of the LVA rats was not as deficient as in this study and was in a range allowing a hormetic response of the organism. Indeed, by looking at the results obtained in the offspring of these rats which were even more deficient in VA, we found that *Lrat* expression was completely inhibited (results not shown and which will be presented in the next publication). It has been suggested that U-shaped dose-response of gene expression can arise when the exogenous ligand is an agonist, which is the case of retinoic acid, and when there is receptor homodimerization, which is the case of RAR. [56] It has also been suggested that U-shaped dose-response of gene expression can arise when specific properties of nuclear receptor and ligands are gathered, which could be the case for RAR and its ligand retinoic acid. The variation in *Lrat* expression, which is expressed dose-dependently across a wide range of dietary VA concentration [55], and which may be considered a target gene for retinoic acid just like CYP26, supports the hypothesis that this U-shaped response of gene expression could be due, directly for VA-responsive genes and indirectly for other genes, to these mechanisms.

Concerning *Isx*, it is interesting to note that it appears to respond increasingly to dietary VA concentration, although the differences between the 3 groups were not significant. This gene encodes for a transcription factor that is positively regulated by retinoic acid. It is assumed that the higher the VA intake, the higher the enterocyte concentration of retinoic acid and the higher *Isx* expression. Furthermore, ISX represses the expression of *Bco1* and of *Scarb1*. [21, 22] Yet we did not observe an inverse relationship between the expression of *Isx* and the expression of *Bco1* and *Scarb1*. Indeed, while the expression of *Isx* tended to increase when the VA content of the diet increased, the expression of intestinal *Bco1* and *Scarb1* tended to develop a U-shaped curve, as in the liver. We have two hypotheses to explain this result in apparent contradiction with the literature. The first one is that the amplitude of the variation in dietary intake of VA, and consequently the variation in the amplitude of the expression of *Isx*, was not sufficiently large to induce significant effects on the expression of *Bco1* and *Scarb1*. The second is that this regulation of the expression of these two genes via ISX, which has been demonstrated only in males, does not exist in females. But these two mechanisms could also coexist.

In conclusion, this study provides new data on hepatic and intestinal metabolism of VA that merit further investigations to assess their pathophysiological consequences. More precisely, we have observed that the bioavailability of β-carotene was much more important in females than in males and, conversely, that β-carotene conversion efficiency was lower in females than in males. This could explain why the blood concentrations of provitamin A carotenoids are generally higher in women than in men. [27, 43] We also observed, in female rats, an unexpected dose response effect of dietary VA concentration on the expression of genes involved in VA metabolism, i.e. a U-shaped dose response, and especially in the liver. We hypothesize that an hormetic regulation of VA metabolism exists, at least in female rats, in order to adapt for variations in dietary VA content. Obviously, this original hypothesis must be confirmed and studied in male rats as well as in other species.

**Acknowledgments**:

The authors thank Charlène Couturier and Lourdes Mounien for technical help and Benjamin Guillet for giving us access to the animal facility.

**PB:** Conceptualization, Methodology, Formal analysis, Resources, Writing - Original Draft, Visualization, Supervision, Project administration, Funding acquisition. **RT:** HPLC analysis, gene expression analysis, figures, statistics. **MD:** HPLC and gene expression analysis. **CH:** HPLC analysis, β-carotene rich emulsion preparation, tissue sampling. **MN:** gene expression analysis, protein quantification, tissue sampling. **PG:** nutritional intervention on rats, rat gavages, blood sampling. **MM:** tissue sampling. **MG:** conceptualization, methodology, nutritional intervention on rats, rat gavages, blood sampling. **ER:** conceptualization, tissue sampling, review and editing. **JFL:** conceptualization, bioinformatic, gene expression validation, writing – review and editing.

**Conflict of interest:**

None of the authors reported a potential competing interest.

**Supporting information:**

This project received funding from both the AlimH department of INRA (ANSSD 2016) and from the G.L.N (Groupe Lipides et Nutrition) in 2017.

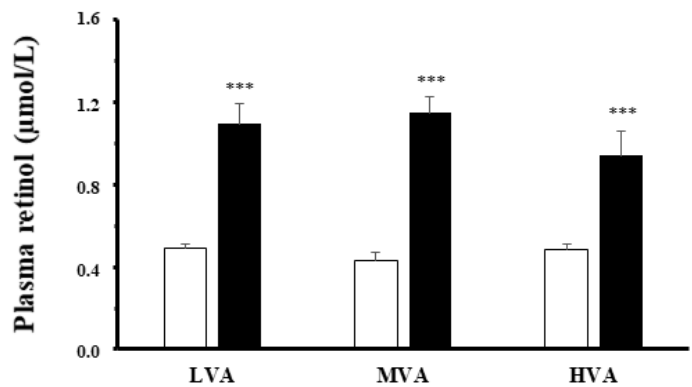

**Figure 1: Plasma retinol concentrations of groups of rats that were fed with diets containing different concentrations of VA for 31-33 weeks.** White bars: females, black bars: males. (**LVA**), low VA diet; (**MVA**), medium VA diet; (**HVA**), high VA diet. Bars represent means ± SEM (n = 6 for females and n = 4 for males). Two-way ANOVA, with the type of diet and the sex of animals as factors, showed a significant effect of sex ($p < 0.0001$) but no effect of diet on plasma retinol concentrations. Asterisks above black bars indicate significant ($p < 0.001$) differences between males and females that ate the same diets.

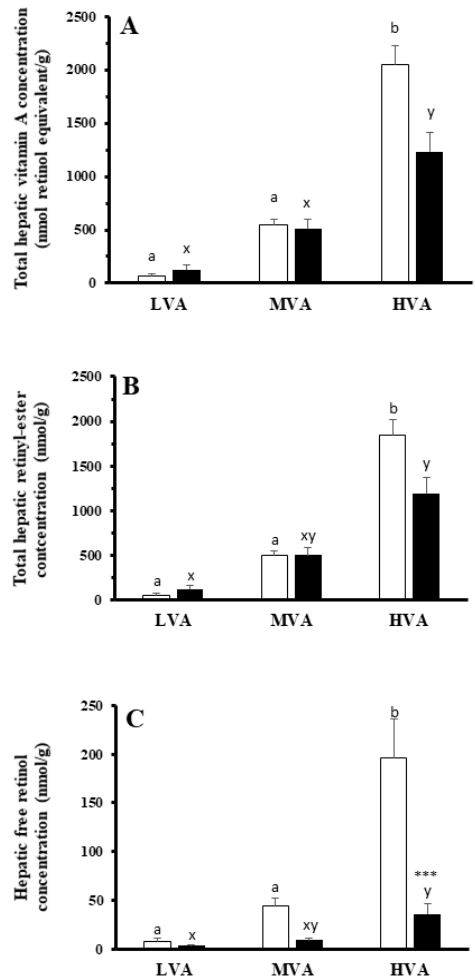

**Figure 2: Hepatic VA concentrations of groups of rats that were fed with diets containing different concentrations of VA for 31-33 weeks. Figure 2A:** total hepatic VA concentrations, i.e. sum of hepatic free retinol and retinyl ester concentrations. **Figure 2B:** total hepatic retinyl ester concentrations, i.e. sum of retinyl palmitate + retinyl stearate + retinyl oleate + retinyl linoleate. **Figure 2C:** hepatic free retinol concentrations. White bars: females, black bars: males. **(LVA),** low VA diet; (**MVA**), medium VA diet; (**HVA**), high VA diet. Bars represent means ± SEM (n = 6 for females and n = 4 for males). Two-way ANOVAs followed by post-hoc Tukey-Kramer tests were performed to compare the effect of diet and sex. Different letters in the a, b, c series indicate significant differences ($p < 0.05$) between groups of females that ate different VA containing diets. Different letters in the x, y, z series indicate significant differences ($p < 0.05$) between groups of males that ate different VA containing diets. Finally, the 3 asterisks on the group of HVA males in figure 2C indicate that the mean of this group is significantly different ($p < 0.001$) from the mean of the corresponding HVA female group.

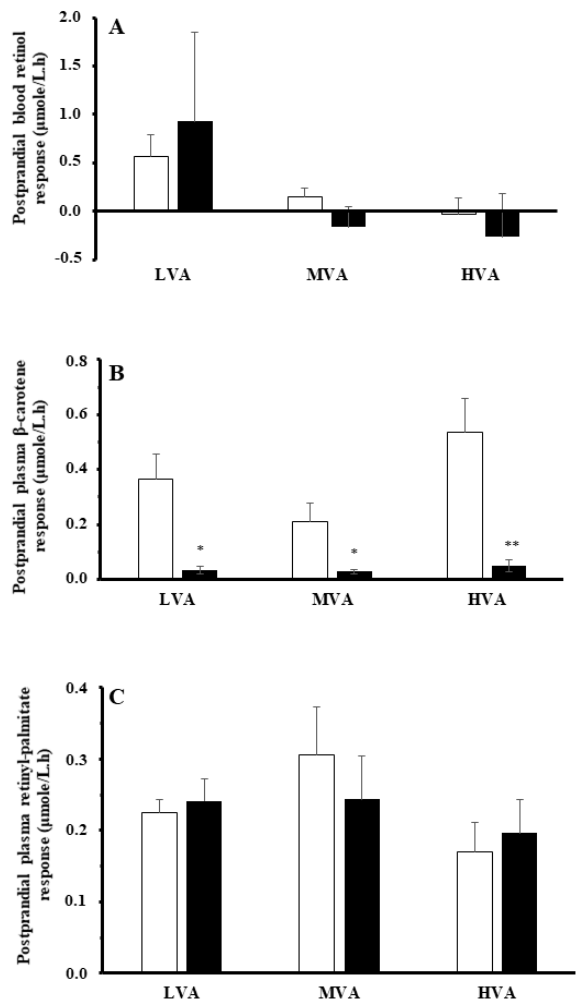

**Figure 3: Postprandial plasma responses of β-carotene and VA after the different groups of rats were force-fed with a high dose of β-carotene.** White bars: females, black bars: males. **(LVA),** low VA diet; **(MVA)**, medium VA diet; **(HVA)**, high VA diet. **Figure 3A:** Plasma retinol responses, i.e. incremental area under the curve (AUC) of the plasma concentrations of retinol measured at regular time intervals up to 6.5 hours after the force-feeding. **Figure 3B:** Plasma β-carotene responses, i.e. AUC of the plasma concentrations of β-carotene measured at the same time intervals as mentioned above. **Figure 3C:** Plasma retinyl palmitate responses, i.e. AUC of the plasma concentrations of retinyl palmitate measured at the same time intervals as mentioned above. Bars represent means ± SEM (n = 6 for females and n = 4 for males). Two-way ANOVAs followed by post-hoc Tukey-Kramer tests were performed to assess the effect of the diet and of the sex of animals. Asterisk(s) indicate(s) that

the means of these groups of males were significantly different from the corresponding group of females (* = $p < 0.05$; ** = $p < 0.01$).

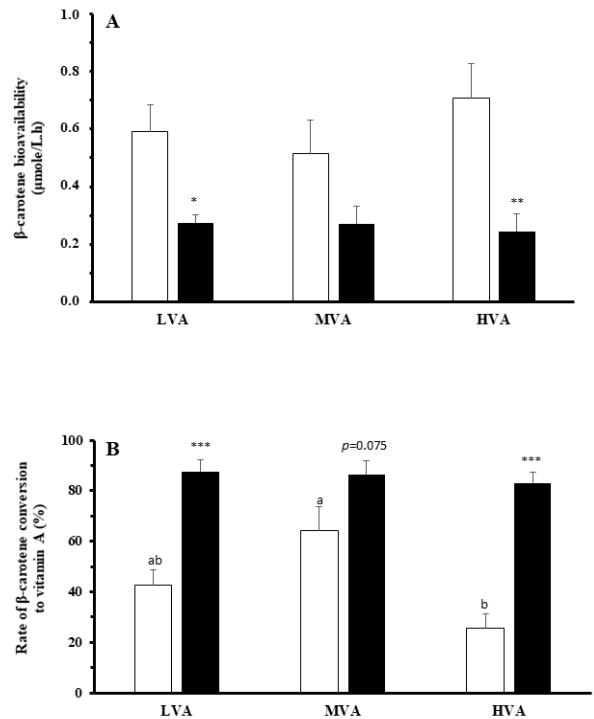

**Figure 4: Effect of the VA content of the diet on the bioavailability of β-carotene and on its conversion efficiency to VA in the intestine. Figure 4A:** β-Carotene bioavailability. This was calculated by adding the postprandial responses to β-carotene and retinyl palmitate because, under these experimental conditions, we assume that the latter originated only from the intestinal metabolism of β-carotene. **Figure 4B:** β-Carotene conversion efficiency to VA. This was estimated by calculating the percentage of bioavailable β-carotene found in the form of retinyl palmitate, i.e. retinyl palmitate AUC / (β-carotene AUC + retinyl palmitate AUC) x 100. White bars: females, black bars: males. **(LVA),** low VA diet; (**MVA**), medium VA diet; (**HVA**), high VA diet. Bars represent means ± SEM (n = 6 for females and n = 4 for males). Two-way ANOVAs followed by post-hoc Tukey-Kramer tests were performed to compare the effect of the diet and the sex of animals. Asterisks indicate that the mean of this group of males was significantly different from the mean of the corresponding group of females (* =

$p < 0.05$; ** = $p < 0.01$; *** = $p < 0.001$). Different letters indicate significant differences ($p < 0.05$) between groups of the same sex that ate different VA containing diets.

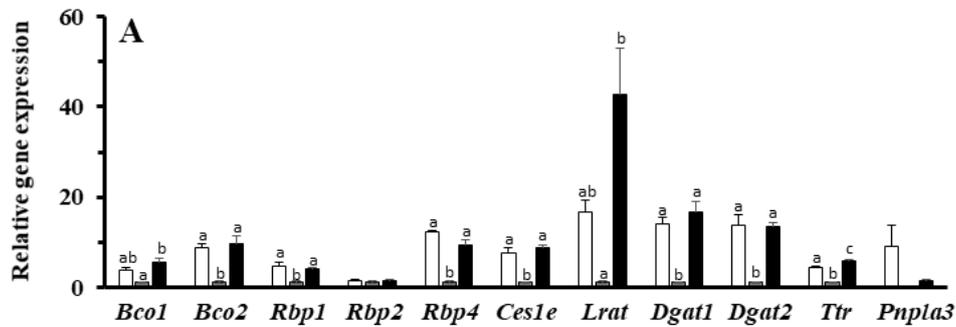

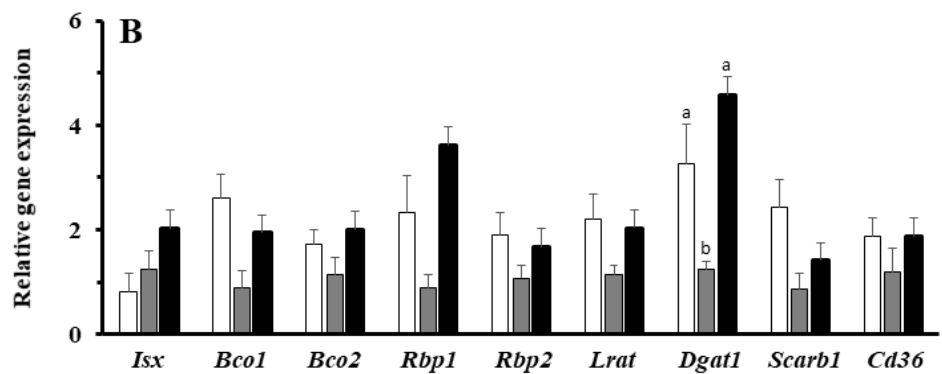

**Figure 5: Effect of the VA content of the diet on the expression of genes involved in VA metabolism in female rats. Figure 5A:** Hepatic genes. **Figure 5B:** Duodenal genes. **(white bars),** low VA diet; (**grey bars**), medium VA diet; (**black bars**), high VA diet. Bars represent means ± SEM (n = 6 for females and n = 4 for males). One-way ANOVA followed by post-hoc Tukey-Kramer tests were performed to compare the effect of the diet on the expression of each gene in each organ. For each gene different letters on the bars indicate significant differences between groups ($p < 0.05$).

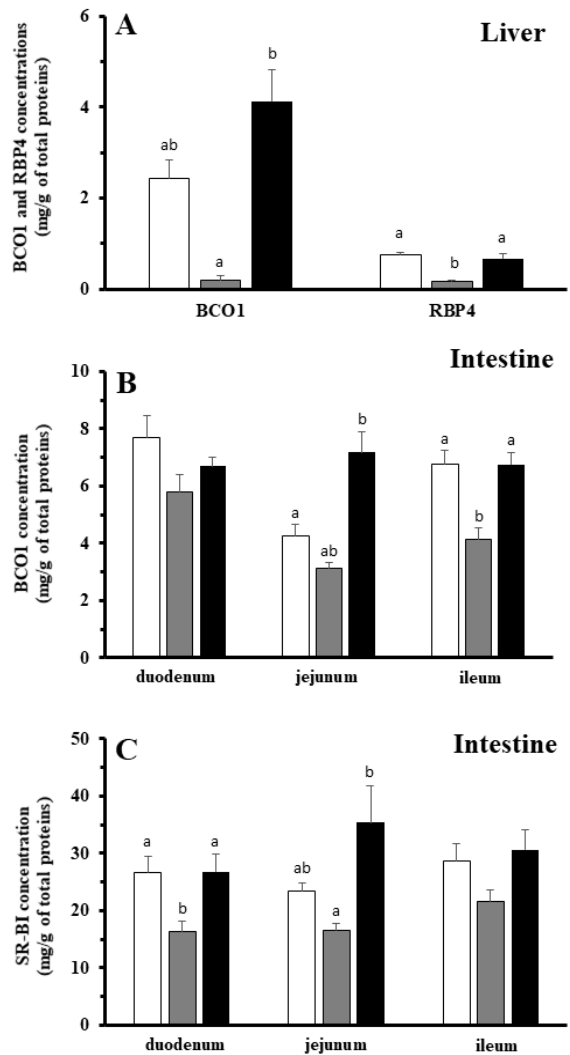

**Figure 6: Effect of the VA content of the diet on the concentration of proteins involved in hepatic and intestinal metabolism of VA in female rats. Figure 6A:** BCO1 and RBP4 concentrations in the liver. Note that BCO1 is involved in cleavage of β-carotene to retinal and RBP4 is responsible, in association with TTR, for the blood transport of retinol. **Figure 6B:** BCO1 concentrations in duodenum, jejunum and ileum. **Figure 6C:** SR-BI concentrations in duodenum, jejunum and ileum. Note that SR-BI is involved in uptake of β-carotene by intestinal cells. **(white bars),** low VA diet; (**grey bars**), medium VA diet; (**black bars**), high VA diet. Bars represent means ± SEM (n = 6 for females and n = 4 for males). One-way ANOVAs followed by post-hoc Tukey-Kramer tests

were performed to compare the effect of the diet on the expression of each protein in each tissue. For each protein different letters on the bars indicate significant differences between groups ($p < 0.05$).

Table 1: Hepatic concentrations of retinyl esters (nmol/g fresh liver) in the different groups of rats.

| Retinyl ester species | LVA | | MVA | | HVA | |
|---|---|---|---|---|---|---|
| | Females | Males | Females | Males | Females | Males |
| Retinyl palmitate | 30 ± 11 **a** | 93 ± 28 **x** | 290 ± 32 **a** | 390 ± 65 **xy** | 1140 ± 128 **b** | 890 ± 148 **y** |
| | (55%) | (75%) | (57%) | (77%) | (62%) | (75%) |
| Retinyl stearate | 16 ± 6 **a** | 16 ± 5 **x** | 153 ± 9 **b** | 65 ± 8 **y\*\*\*** | 466 ± 35 **c** | 168 ± 25 **y\*\*\*** |
| | (29%) | (13%) | (30%) | (13%) | (25%) | (14%) |
| Retinyl oleate | 5 ± 2 **a** | 9 ± 3 **x\*** | 34 ± 3 **a** | 32 ± 4 **x** | 127 ± 15 **b** | 77 ± 9 **y\*** |
| | (9%) | (7%) | (7%) | (6%) | (7%) | (6%) |
| Retinyl linoleate | 4 ± 2 **a** | 6 ± 2 **x** | 29 ± 3 **a** | 19 ± 2 **x** | 117 ± 15 **b** | 60 ± 10 **y\*** |
| | (7%) | (5%) | (6%) | (4%) | (6%) | (5%) |

(**LVA**), low VA diet; (**MVA**), medium VA diet; (**HVA**), high VA diet. Means ± SEM (n = 6 for females and n = 4 for males). The% numbers represent the percentages of each retinyl ester in each group. Two-way ANOVAs followed by post-hoc Tukey-Kramer tests were performed to compare the effect of the diet and the sex of animals on the concentration of each retinyl ester species. Different letters indicate significant differences ($p < 0.05$) between groups of the same sex that ate different VA containing diets. Asterisks indicate that the mean of this group of males was significantly different from the mean of the corresponding group of females (* = $p < 0.05$; *** = $p < 0.001$).